\documentclass[preprints,article,accept,moreauthors,pdftex]{Definitions/nomdpi} 

\datereceived{} 
\pdfoutput=1

\usepackage{graphicx}
\usepackage{booktabs} 
\usepackage{subcaption} 
\usepackage{multirow} 
\usepackage{rotating} 

\Title{rWISDM: Repaired WISDM, a Public Dataset for Human Activity Recognition}

\Author{Mohammadreza Heydarian $^{1,2}$\orcidA{} and Thomas E. Doyle $^{1,3,4,}$*\orcidB{}}

\address{%
$^{1}$ \quad Department of Electrical and Computer Engineering, McMaster University, Hamilton, ON, Canada\\
$^{2}$ \quad Department of Computer Science, Vali-e-Asr University of Rafsanjan, Rafsanjan, Kerman, Iran\\
$^{3}$ \quad School of Biomedical Engineering, McMaster University, Hamilton, ON, Canada\\
$^{4}$ \quad Vector Institute for Artificial Intelligence, Toronto, ON, Canada}

\corres{Correspondence: doylet@mcmaster.ca; www.biomedic.ai}

\abstract{Human Activity Recognition (HAR) has become a spotlight in recent scientific research because of its applications in various domains such as healthcare, athletic competitions, smart cities, and smart home. While researchers focus on the methodology of processing data, users wonder if the Artificial Intelligence (AI) methods used for HAR can be trusted. Trust depends mainly on the reliability or robustness of the system. To investigate the robustness of HAR systems, we analyzed several suitable current public datasets and selected WISDM for our investigation of Deep Learning approaches. While the published specification of WISDM matched our fundamental requirements (e.g., large, balanced, multi-hardware), several hidden issues were found in the course of our analysis. These issues reduce the performance and the overall trust of the classifier. By identifying the problems and repairing the dataset, the performance of the classifier was increased.  This paper presents the methods by which other researchers may identify and correct similar problems in public datasets. By fixing the issues dataset veracity is improved, which increases the overall trust in the trained HAR system.}


\begin{document}


\section{Introduction}
\label{intro}

Recognition of human activity has broad applications in various domains such as healthcare \cite{health_review2019,health_review2021,health_trends2021}, athletic competitions \cite{DS_vis_athl_FineGym2020}, smart cities \cite{dRNN_cities_phone2021}, and smart home \cite{DS_ext_VanKasteren2010,DS_ext_CASAS2013}. Different data sources could be used to apply Human Activity Recognition (HAR). Data sources could be categorized as vision-based \cite{vision_review2020,vision_sensor_review2020,vision_review2021} and sensor-based devices, whereas the latter could be classified as external sensors \cite{DS_ext_VanKasteren2010,DS_ext_CASAS2013} and internal sensors \cite{DS_int&ext_OPPORTUNITY2010,DS_UCI_HAR2013,DS_WISDM2019}. External sensors, similar to vision-based sensors, are installed in different locations of the environment. Internal sensors - embedded in the wearable garments/straps or the devices such as smartwatches and smartphones - are attached to the human body. While vision-based and external sensors are mainly used for smart cities and smart home, internal sensors are used for healthcare and athletic applications.

In the recent decade, most people have used smartphones \cite{phone_trends2021} and carry them wherever they go most of the time. With some embedded sensors and the computation ability of a smartphone, it became a popular target device to be considered for new applications other than its original purpose as a smartphone. Accelerometer, gyroscope, magnetometer, and GPS sensors can record the smartphone’s movement, which could be used to train a classifier for recognizing human activities.

Human activities could be categorized into single and complex activities \cite{DS_int&ext_OPPORTUNITY2010,DL_review2018,dML_complex2020}. Single activities are divided into moving, static, and dynamic activities; moving activities such as walking, upstairs, downstairs, jogging, running, typing, cooking, etc.; static activities such as standing, sitting, lying; and dynamic activities such as standing up from sitting/lying, sitting from standing/lying, stepping in/out a car, etc., which usually are referred to as transition or action. Complex activities are divided into concurrent and interleaved activities; concurrent activities are two activities that are observed concurrently through a recorded activity sequence; interleaved activity refers to the situation that one activity is paused then resumes after executing another activity. Some of the HAR datasets contain falls activity for the research on concussion \cite{health_review2019}. 

Creating a human activity dataset is a challenging task that needs time, money, and effort. De-La-Hoz-Franco \textit{et al.} \cite{DS-review-2018} provided a systematic review for sensor-based HAR datasets that were created until 2018. Bian \textit{et al.} \cite{sensor_review2022} presented a survey on the state-of-the-art sensing modalities in HAR tasks. Fortunately, many researchers who created the datasets made them publicly available. For example, UCI-HAR \cite{DS_UCI_HAR2013}, MobiAct \cite{DS_MobiAct2016}, UniMiB SHAR \cite{DS_UniMib2017}, and WISDM \cite{DS_WISDM2019} are publicly available for research on HAR. Table \ref{tab:four_datasets} shows the properties of these four public datasets. The WISDM dataset was published in 2019 under the HAR datasets but was originally used for the user authentication \cite{DS_WISDM2019} and reported in an M.Sc.\ thesis for HAR application. The published specification of the WISDM dataset with the 51 subjects and 17 activities recorded using accelerometer and gyroscope of smartphone and smartwatch matched our fundamental requirements (e.g., large, balanced, multi-hardware). 

The creator of WISDM reported having fixed the smartphone axes' orientation and the 20Hz sampling rates. However, we found that the smartphone orientation was not consistent for all subjects, and the sampling frequency was not constant for all subjects and activities. To use this dataset for HAR, we repaired it for consistency in extracted features. Based on repairing this dataset, our contribution is listed as below:
\begin{itemize}
\item Identifying and repairing the different orientations of smartphones for each subject and activity
\item Identifying and repairing the different sampling frequencies for each subject and activity
\item Using WISDM dataset for HAR where originally it was designed for user authentication
\end{itemize}

This paper is organized as follows. Section 2 briefly refers to some of the publicly available HAR datasets. Section 3 describes our proposed method for repairing the smartphone orientation and repairing the recording frequency inconsistency. Section 4 shows the result of repairing the dataset visually and statistically, followed by a discussion section. The last section concludes the contribution of this paper to the HAR applications with suggestions for future work.

\section{Background}
\label{background}
Human Activity Recognition (HAR) has become an increasingly important research area given the now ubiquitous nature of smartphones \cite{phone_trends2021}. However, it was under investigation before the invention of smartphones, using embedded sensors in wearable garments or straps \cite{health_review2019}. HAR has broad applications in various domains such as healthcare, athletics, smart cities, smart home, human-computer interface, cyber-physical system, etc. Each of these applications needs a specific type of dataset with some shared attributes. Some of these datasets could be used for more than one application. For example, the HAR dataset for recognizing activities such as walking and running could be used for user authentication applications too \cite{DS_WISDM2019}. 

Machine learning methods that are used for these applications have a significant overlap \cite{ML_review2020}. Although, in the past, primarily conventional machine learning methods such as Support Vector Machine (SVM), Random Forrest (RF), and Neural Network (NN) were used for HAR, in recent years, with the availability of the larger datasets and powerful High-Performance Computing (HPC) systems, the Deep Learning (DL) models are used by researchers and industries \cite{DL_review2018,DL_review2021,DL_review2022}.  

Since there are broad applications of HAR, and because some researchers look through the applications from a different perspective, they have to create their datasets. Therefore, we see many HAR datasets that many of which are publicly available. Although many datasets are publicly available, the lack of standardization makes them useless for other researchers and makes it impossible to compare methods on different datasets. Consequently, we see the lack of suitable publicly available datasets for new researchers to explore the field of HAR \cite{health_trends2021}. 

Before the popularity of smartphones, the external sensors and the embedded sensors in wearable garments/straps were used for creating the HAR datasets. For example, Van Kasteren \textit{et al.} \cite{DS_ext_VanKasteren2010} created a dataset known as the VANKASTEREN dataset for investigating the smart home by installing sensors in one apartment with three rooms and one house with six rooms. They recorded data for 25 days at the apartment where a 26 years-old man was living. They recorded data for 19 days at the house where a 57 years-old man was living. For another example, Cook \textit{et al.} \cite{DS_ext_CASAS2013} created CASAS dataset as a smart home in a box. They collected data from 32 smart homes, 19 with a single resident, four with two residents, and the rest with larger families or residents with pets. In another research, Roggen \textit{et al.} \cite{DS_int&ext_OPPORTUNITY2010} used a mixture of external and internal sensors to create the OPPORTUNITY dataset from 12 subjects for recognition of complex human activities. 

Some recent datasets were created using wearable sensors. For example, Bhat \textit{et al.} \cite{DS_int_wHAR2020} created a dataset using wearable sensor with 22 participants for seven activities. Logacjov \textit{et al.} \cite{DS_int_HARTH_2021} created a dataset from inertial and wearable stretch sensors and a chest-mounted camera with 22 participants for 12 activities. Khan \textit{et al.} \cite{DS_int_kinect2022} created a dataset using Kinect V2 sensor with 20 participants for 12 activities. Li \textit{et al.} \cite{DS_int_homemade2022} created a dataset by locating a sensor below the knee with five participants for six activities. 

With the availability and popularity of smartphones \cite{phone_trends2021}, more HAR datasets based on its embedded accelerometer or gyroscope are available \cite{DS-review-2018}. Most of these datasets collected data from a few subjects. For example, Reiss \textit{at al.} \cite{DS_PAMAP2012} created PAMAP2 dataset based on collecting data from nine participants. Garcia-Gonzalez \textit{et al.} \cite{DS_reallife2020} created a real-life-HAR-dataset based on the accelerometer, gyroscope, magnetometer, and GPS of the smartphone from 19 participants. 

Some HAR datasets were created based on more than 30 participants to collect data. For example, Anguita \textit{et al.} \cite{DS_UCI_HAR2013} created the UCI-HAR dataset based on collected data from 30 participants using a smartphone accelerometer. The UCI-HAR has been used as one of the benchmarks to compare HAR methods. Vavoulas \textit{et al.} \cite{DS_MobiAct2016} created MobiAct dataset based on collected data from 57 participants using smartphone accelerometer. Micucci \textit{et al.} \cite{DS_UniMib2017} created UniMib dataset based on collected data from 30 participants using smartphone accelerometer. Weiss \textit{et al.} \cite{DS_WISDM2019} created WISDM dataset based on collected data from 51 participants using smartphone/smartwatch accelerometer/gyroscope. Table \ref{tab:four_datasets} shows the specification of four public datasets with a relatively high number of subjects.

\begin{table}[ht]
	\begin{adjustwidth}{-\extralength}{0cm}
    \centering
    \caption{Comparing four public datasets. The cells with the $``-"$ sign means there is no information available for those cells}
    \label{tab:four_datasets}
    \resizebox{0.95\linewidth}{!}{ 
\begin{tabular}{|c|c|c|c|c|c|c|c|c|c|c|c|c|c|c|} 
\hline
\multicolumn{1}{|c|}{\multirow{2}{*}{Dataset}} & \multicolumn{1}{c|}{\multirow{2}{*}{Year}} 
& \multicolumn{1}{c|}{\multirow{2}{*}{\begin{tabular}[c]{@{}c@{}}No. of\\ Subj.\end{tabular}}} 
& \multicolumn{1}{c|}{\multirow{2}{*}{\begin{tabular}[c]{@{}c@{}}Act. Len.\\ (sec)\end{tabular}}} 
& \multicolumn{1}{c|}{\multirow{2}{*}{\begin{tabular}[c]{@{}c@{}}Freq.\\ (Hz)\end{tabular}}} 
& \multicolumn{1}{c|}{\multirow{2}{*}{\begin{tabular}[c]{@{}c@{}}Raw\\ Data\end{tabular}}} 
& \multicolumn{2}{c|}{Phone} & \multicolumn{2}{c|}{Watch} & \multicolumn{2}{c|}{Gender} 
& \multicolumn{1}{c|}{\multirow{2}{*}{\begin{tabular}[c]{@{}c@{}}Age\\ \{avg,std\}\end{tabular}}} 
& \multicolumn{1}{c|}{\multirow{2}{*}{\begin{tabular}[c]{@{}c@{}}Height(cm)\\ \{avg,std\}\end{tabular}}} 
& \multicolumn{1}{c|}{\multirow{2}{*}{\begin{tabular}[c]{@{}c@{}}Weight(Kg)\\ \{avg,std\}\end{tabular}}} \\ 
\cline{7-12} \multicolumn{1}{|c|}{} & \multicolumn{1}{c|}{} & \multicolumn{1}{c|}{} & \multicolumn{1}{c|}{} 
& \multicolumn{1}{c|}{} & \multicolumn{1}{c|}{} & \multicolumn{1}{c|}{acc} & \multicolumn{1}{c|}{gyr} & \multicolumn{1}{c|}{acc} 
& \multicolumn{1}{c|}{gyr} & \multicolumn{1}{c|}{F} & \multicolumn{1}{c|}{M} & \multicolumn{1}{c|}{}   
& \multicolumn{1}{c|}{} & \multicolumn{1}{c|}{}

\\ \hline
UCI HAR & 2012 & 30 & 30-36 & 50 & N & Y & N & N & N  & - & -  & \begin{tabular}[c]{@{}c@{}}19-48\\ (-,-)\end{tabular} & -  & -  

\\ \hline
MobiAct & 2016 & 57 & 10-300 & 20 & Y  & Y & N & N & N & 15 & 42 & \begin{tabular}[c]{@{}c@{}}20-47\\ (25,4)\end{tabular}   
& \begin{tabular}[c]{@{}c@{}}160-193\\ \{175,4\}\end{tabular} & \begin{tabular}[c]{@{}c@{}}50-120\\ \{76.6,14.4\}\end{tabular}     

\\ \hline
UniMiB SHAR & 2016 & 30   & 15-30 & 50 & Y  & Y & N & N & N & 24 & 6  & \begin{tabular}[c]{@{}c@{}}18-60\\ (27,11)\end{tabular}  
& \begin{tabular}[c]{@{}c@{}}160-190\\ \{169,7\}\end{tabular} & \begin{tabular}[c]{@{}c@{}}50-82\\ \{64.4,9.7\}\end{tabular}       

\\ \hline
WISDM   & 2018 & 51 & 180 & 20 & Y & Y & Y & Y & Y & - & - & \begin{tabular}[c]{@{}c@{}}18-25\\ (-,-)\end{tabular} & -  & -  

\\ \hline
\end{tabular}
} 
	\end{adjustwidth}
\end{table}

There is an increasing amount of research on methods for recognizing human activities. In some works, researchers needed to create their dataset based on their goal. In this research, we neither want to investigate the HAR methods for better accuracy nor wish to create a new HAR dataset. Our goal was to examine the robustness of HAR methods and research on measuring trust in Artificial Intelligence (AI) for HAR applications. Thus, we chose one of the existing public datasets and implemented a classifier from the literature. In the next section, we explain how we chose a public HAR dataset and a classifier, followed by explaining our contributions in repairing the selected dataset.   

\section{Materials and methods}
\label{methods}
Our experience shows that not all public datasets contain raw data and/or meta-data such as participants’ demographics (e.g., age, weight, etc.). For example, the cells with the $``-"$ sign in Table \ref{tab:four_datasets} means there is no information available for those cells, where the column $``Raw Data"$ in this Table shows which dataset contains the raw data. Also, for several datasets, we found missing data (e.g., one or more activities of some participants). Like many research groups that planned to collect our own data, the pandemic has interrupted our human activity data collection. Consequently, public human activity datasets have become very popular for continuing such research.

In most systems, particularly for the HAR system, which is implemented on the data from humans, the training dataset may cause bias(es) because of the differences in subject behaviours. In other words, the behaviour of one subject could be different from other subjects (e.g., one's walking may be similar to another’s stair) based on age, gender, weight, height, BMI, culture, race, athletic background, etc. With consideration towards trust and explainability of AI systems, more detail on subject attributes is necessary to help reduce bias and increase the fairness of AI models. To improve the robustness of our AI, human activity datasets need to increase the number of participants, the diversity of participants, the veracity of the data, and the hardware variation for the same purposed sensors (i.e., different devices with accelerometers). 

With the current human activity data collection restrictions (i.e., the pandemic), publicly available HAR datasets were assessed for our machine learning requirements. We believe these requirements will be similar to other research groups working to develop various AI applications related to human activity. Section \ref{public_data} explains how the assessment was conducted and presents the rationale for the selection of one of those datasets.

\subsection{Public datasets}
\label{public_data}
Table \ref{tab:four_datasets} shows the specification of four public datasets with a relatively high number of subjects when compared to the other available datasets. For example, there are many other datasets for HAR with less than 30 subjects, some of them with only five or even one subject (i.e., reviewed by \cite{DS_MobiAct2016,sensor_review2019}.) 
Regarding our goals, we were looking for a dataset to investigate the robustness of machine learning, especially a deep Convolutional Neural Network (dCNN). Therefore, we searched for a dataset with many subjects and a more significant length of each recorded activity. 
Also, we considered finding a dataset that recorded human activity using different smartphone devices and other devices such as a smartwatch with the same sensor (i.e., accelerometer). Initially, the WISDM dataset \cite{DS_WISDM2019} was a promising dataset that satisfied most of our requirements, although the subjects’ metadata such as gender, height, and weight were missing. The WISDM dataset contains data from 51 subjects coded 1600 through 1650, with 17 activities coded from A through S (with missing N). The data was collected using an embedded accelerometer, gyroscope, and magnetometer of smartphone and smartwatch. In this study, we only focus on five activities (i.e., walking (A), Jogging (B), stairs (C), sitting (D), and standing (E)) recorded by the accelerometer of smartphone and smartwatch. 

\subsection{Identifying and repairing smartphone axes orientation}
\label{rep_ori}
The protocol for creating the WISDM dataset \cite{DS_WISDM2019} states that the axes orientation and the location of the smartphones were the same for all subjects. However, our investigation of recorded smartphone-accelerometer signals shows that the smartphone's orientation was different for most of the subjects from what was reported. This could result from putting the smartphone in the subject's pants packet with the incorrect orientation, or the the smartphone's position was altered inadvertently after it was placed in the subject’s pants pocket, which was not a secure holder for the smartphone. 

\begin{figure} [t]
\begin{adjustwidth}{-\extralength}{0cm}

\centering
\begin{subfigure}{.48\linewidth}
  \centering
  \includegraphics[width=.99\linewidth]{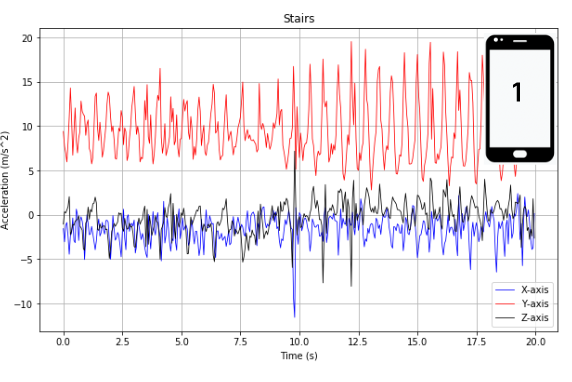}
  \caption{Case 1}
  \label{fig:orient_1}
\end{subfigure}%
\hspace{2mm}
\begin{subfigure}{.48\linewidth}
  \centering
  \includegraphics[width=.99\linewidth]{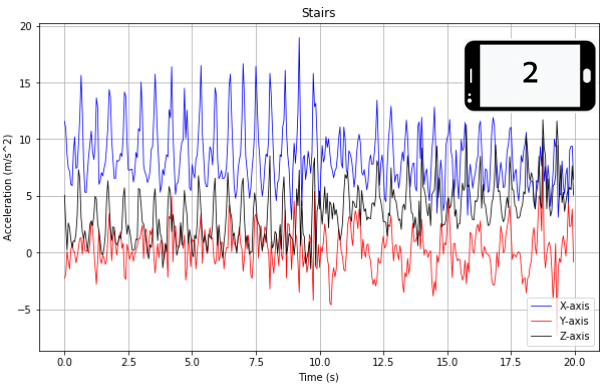}
  \caption{Case 2}
  \label{fig:orient_2}
\end{subfigure}%
\vspace{3mm}
\begin{subfigure}{.48\linewidth}
  \centering
  \includegraphics[width=.99\linewidth]{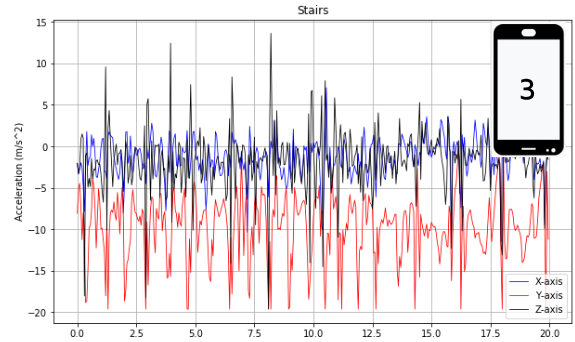}
  \caption{Case 3}
  \label{fig:orient_3}
\end{subfigure}%
\hspace{2mm}
\begin{subfigure}{.48\linewidth}
  \centering
  \includegraphics[width=.99\linewidth]{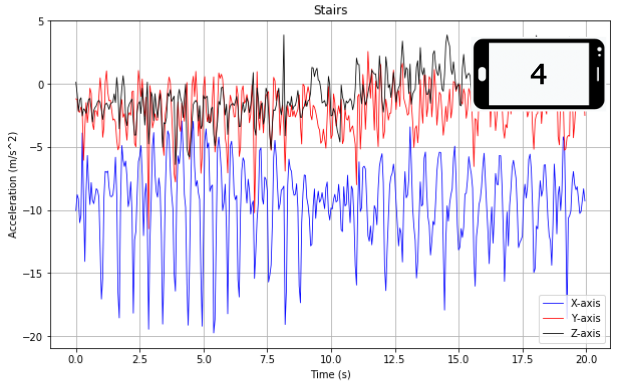}
  \caption{Case 4}
  \label{fig:orient_4}
\end{subfigure}%
\caption{Recorded signals from smartphone accelerometer XYZ axes for different orientations of the smartphone. Amplitude of the signal is acceleration ($m/s^2$) over time ($s$). The blue, red, and black show X, Y, and Z axes of accelerometer signals, respectively}
\label{fig:phone_orientations}
\end{adjustwidth}
\end{figure}

The inconsistency in the smartphone axes orientation could be identified by comparing the mean and standard deviation of all signal axes to each other for each activity. This problem could be empirically observed by comparing the recorded signals for each activity of a subject to the same activity signal of different subjects, as shown in Figure \ref{fig:phone_orientations}.

Figure \ref{fig:phone_orientations} illustrates recorded smartphone-accelerometer signals from four different subjects during the stairs activity, each of which with a different orientation of the smartphone. We named these four orientations cases 1 through 4 as shown in Figures \ref{fig:orient_1} through \ref{fig:orient_4}, respectively. For case 1, the Y-axis of the accelerometer is parallel and has the same orientation as the gravity. In contrast, in case 3, the Y-axis is parallel but opposite to the gravity. In case 2, the X-axis is parallel and has the same orientation as the gravity, whereas, for case 4, the X-axis is parallel but opposite to the gravity.

As shown in Figure \ref{fig:phone_orientations}, we only considered the smartphone facing out from the subject. There are four other orientations to be considered when the smartphone faces the subject, which only affects the Z-axis signal to be inverse (i.e., does not affect the X and Y axes signals). Since the smartphone was placed in the subject's pants pocket, in activities such as sitting, the Z-axis would be parallel to the gravity and has the highest amplitude. 

To unify the smartphone axes' orientation, we corrected the signals to match the orientation of case 1. To do this, we shift the signal whose average amplitude is negative, so that the signal will have the same absolute average but positive. Then we exchange X and Y axes if the average amplitude of the X-axis is more significant than the average amplitude of the Y-axis. We move the signals to have the positive amplitude neither by calculating the absolute number nor by multiplying with the negative one (-1) but shift the signals by adding two times the absolute average of the amplitude. 
Not by multiplying with a negative one, because we want to avoid ending with a mirror of the signal (about horizontal axis), which will lose the pattern of the activity. 
Not by calculating the absolute number, because there could be a situation (especially for the axes that are not parallel to gravity) that the signal oscillates to positive and negative amplitudes. For example, in Figure \ref{fig:phone_orientations}, the Z-axis signal is oscillating around zero for all four cases, the X-axis is oscillating around zero for cases 1 and 3, and the Y-axis is oscillating around zero for cases 2 and 4.

Rarely, for some subjects (e.g., subjects 1638 and 1648, where subjects coded from 1600 to 1650), the orientation of the smartphone was changed for a few seconds during some of the activities; thus, we applied the repairing algorithm on segmented windows of the signal to cover this situation.

\subsection{Identifying and repairing sampling frequency}
\label{rep_freq}
From the description document uploaded with the WISDM dataset, we found that the total number of sample points for the smartphone accelerometer is significantly higher than the total number of sample points for the smartphone gyroscope and the smartwatch accelerometer and gyroscope, as shown below:
\begin{itemize}
\item raw/phone/accel: 4,804,403
\item raw/phone/gyro: 3,608,635
\item raw/watch/accel: 3,777,046
\item raw/watch/gyro: 3,440,342
\end{itemize}

The expected size was calculated by multiplying the number of subjects (51), number of activities (17), length of each activity (180s), and the frequency (20Hz), which is equal to 3,121,200. This difference in the total number of sample points motivated us to investigate the length of each activity recorded for each subject. Surprisingly we found that the length of the activities is varied for different subjects. For example, based on the smartphone accelerometer data, the number of sample points for one activity for some subjects is almost 3600 points (e.g., subjects 1600, 1602, 1604, etc.), 4500 points (e.g., subjects 1601, 1647, etc.), 7150 points (e.g., subject 1629), or 8950 points (e.g., subjects 1644, 1635, etc.). In some cases, the length of some activities is different for the same subject (e.g., subject 1602). For example, for subject 1602, each activity is almost 3600 sample points long, but activities H, I, J, and K are nearly 8950, 7950, 8950, and 8100 sample points, respectively. 

Based on the reported recording time of three minutes (i.e., 3600 sample points with 20Hz sampling frequency), there is a low possibility of mistakenly recording an activity longer than planned. For example, recording 8950 sample points which are nearly $7.5$ minutes with a frequency of 20Hz instead of the $3$ minutes recording, is unlikely. Thus, we deeply inspected the raw recorded data and calculated the sampling frequency based on the recorded Unix timestamp for each sample point. The calculated sampling frequency shows that the data was recorded with various sampling frequencies, although it was reported that the sampling frequency was fixed to 20Hz. We believe this is because of using three different smartphones (i.e., Google Nexus 5, Google Nexus 5X, and Samsung Galaxy S5). While converting the Unix time to the calendar time, we notice that the recorded Unix time does not match the recording date (i.e., considering the experiment year was 2017). However, the increment in the timestamp is correct and consistent, which could be used for calculating the sampling frequency. 

The calculated sampling frequency for activities with almost 8950 sampling points is 50Hz. After resampling to 20Hz, the length of the signal was reduced to nearly 3600 sample points, equal to the expected number of samples for three minutes recording with 20Hz sampling frequency. 

Since the smartphone orientation and sampling frequency are independent, the order for repairing the smartphone axes’ orientation and the sampling frequency is not important.

\subsection{Preprocessing data}
\label{prep_data}
After reorienting the XYZ axes and resampling signals, we processed data to train a deep learning method. The recorded signal for each subject is 180 seconds long per activity. To simulate the real-time HAR, we segmented the signals to 5-second windows with a one-second stride for the data augmentation. The minimum window size that gave a good performance for the classifier was a 3-second window. We received better results on a 5-second window than 3- and 10-second windows. 

While observing the signals, we see that the first few seconds of the signals were recorded before the activity started. These first few seconds of the signal are considered transition action (e.g., standing to sitting, etc.) and need to be removed from the activity. Therefore, we removed the first 15 seconds of the signals to avoid the not activity-related part of the signals.   

We split subjects to 41/5/5 out of 51 subjects for Train/Validation/Test sets. This method of splitting data prevents the Train/Validation/Test sets from having overlap on subjects. Later we used the leave-one-out method for reporting the performance of the classifier.

\subsection{Classification methodology}
\label{classifier}
The WISDM dataset was initially used for the user authentication purpose \cite{DS_WISDM2019}. The different orientations of the smartphone and the different frequencies of recorded signals would bias the classification algorithms in increasing the authentication accuracy and decreasing the Human Activity Recognition accuracy. 

Our goal was to use this dataset to investigate the robustness of a deep learning method for HAR. Before discovering the problems with the smartphone axes orientation and the recording frequency, we developed a deep Convolutional Neural Network (dCNN) based on the method presented by Wan \textit{et al.} \cite{dCNN-2020}. We trained the dCNN on the data from the smartphone accelerometer and the smartwatch accelerometer signals. Then, we used the same model after identifying and repairing the problems with the WISDM dataset. 

\begin{figure}[t]
    \centering
    \includegraphics[scale=0.45]{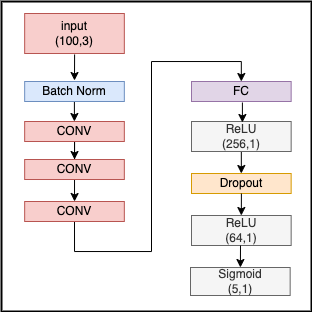}
    \caption{Visualization of the CNN model}
    \label{fig:HAR_model}
\end{figure}

Since the developed model is not the focus of this paper, we only give a brief view of the model. 
The dCNN model contains a Batch-Normalization layer for standardizing the input of size 100 (i.e., five seconds of signal with 20Hz) followed by three CNN blocks and two fully connected layers. The size of the output layer is equal to the number of classes (i.e., five activities such as Walking, Jogging, Stairs, Sitting, and Standing). We added a dropout layer to prevent the model from overfitting. Figure \ref{fig:HAR_model} shows a schematic visualization of the Convolution Neural Network (CNN) that we used. We trained the model for 100 epochs on inputs from 41 randomly selected subjects. We used the 41/5/5 out of 51 subjects for Train/Validation/Test sets to tune the dCNN model. We trained the model on the original and repaired dataset. As the performance of the HAR system is very subjective (e.g., one's walking may be similar to another’s stair), we average the model performance using the leave-one-out method for calculating the performance of the classifier.

\subsection{Hardware cross-validation}
\label{cross_val}
To investigate the model robustness caused by repairing the dataset, we trained the model on the smartphone accelerometer data and tested it on the smartwatch accelerometer data and vice versa. For the cross-validation, we used data from all 51 subjects for training the model on smartphones and tested the trained model on data from all 51 subjects from smartwatches (and vice versa). 

\section{Results}
\label{result}
\begin{figure} [t]
\begin{adjustwidth}{-\extralength}{0cm}
\centering
\begin{subfigure}{.48\linewidth}
  \centering
  \includegraphics[width=.99\linewidth]{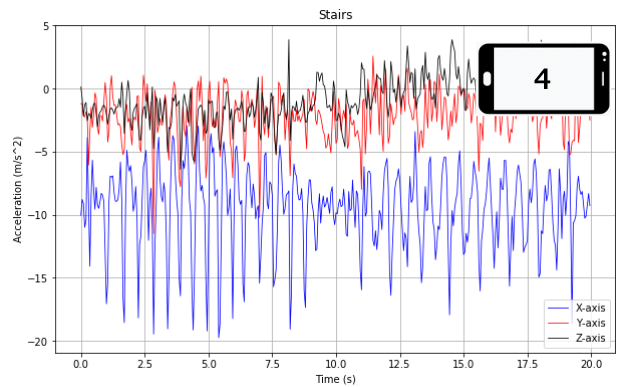}
  \caption{Original signal}
  \label{fig:repair_orient_a}
\end{subfigure}%
\begin{subfigure}{.48\linewidth}
  \centering
  \includegraphics[width=.99\linewidth]{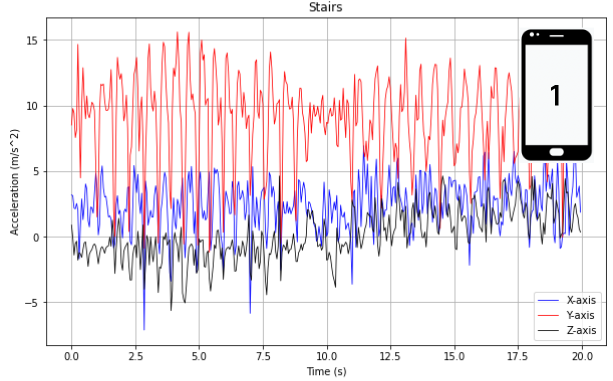}
  \caption{Repaired signal}
  \label{fig:repair_orient_b}
\end{subfigure}
\caption{Signals before and after repairing smartphone axes orientation. The number on the phone shows the case number defined in Figure \ref{fig:phone_orientations}}
\label{fig:repair_orient}
\end{adjustwidth}
\end{figure}

After repairing the smartphone axes orientation, the mean amplitude of the Y-axis would be greater than the mean amplitude of the X-axis, while the mean amplitude of all XYZ axes would be positive.    
Figure \ref{fig:repair_orient} shows one example of repairing the smartphone axes orientation. Figure \ref{fig:repair_orient_a} and Figure \ref{fig:repair_orient_b} show the original and repaired signals, respectively. 

\begin{figure} [t]
\begin{adjustwidth}{-\extralength}{0cm}
\centering
\begin{subfigure}{.68\linewidth}
  \centering
  \includegraphics[width=.99\linewidth]{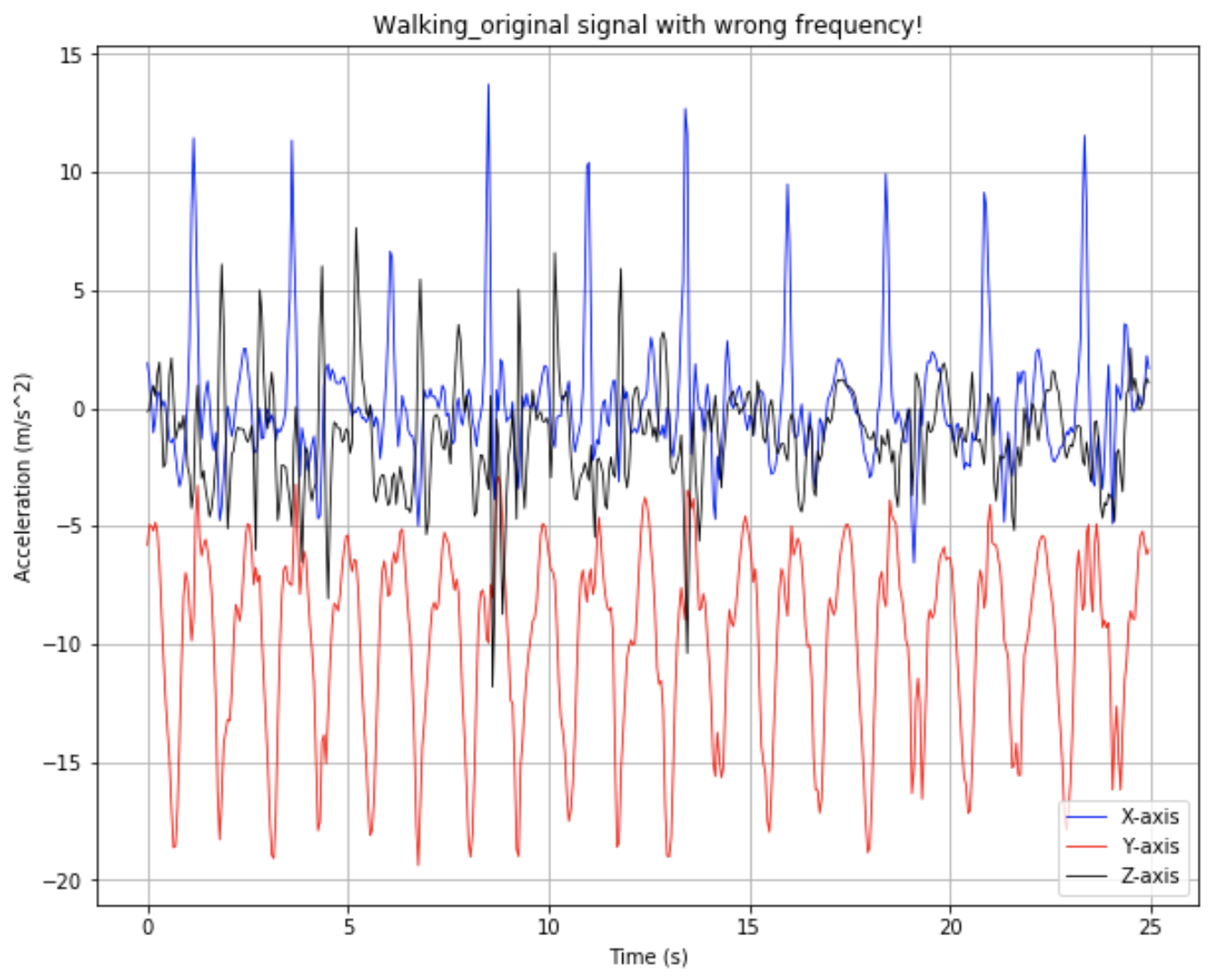}
  \caption{Original signal (50Hz). Time is based on 20Hz}
  \label{fig:repair_freq_1}
\end{subfigure}%
\begin{subfigure}{.31\linewidth}
  \centering
  \includegraphics[width=.99\linewidth]{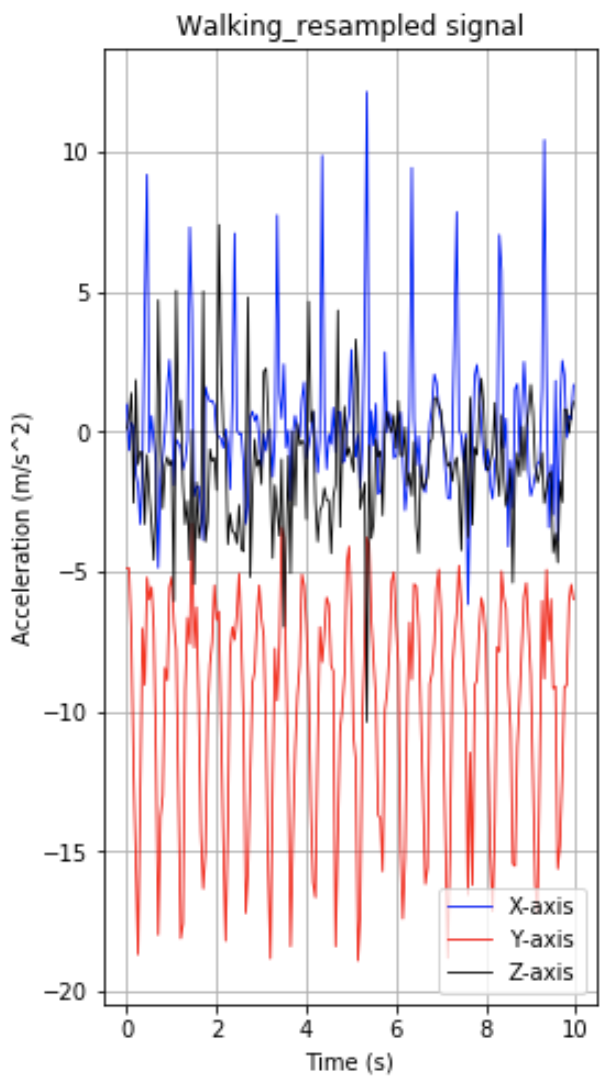}
  \caption{Resampled signal (20Hz)}
  \label{fig:repair_freq_2}
\end{subfigure}
\caption{Signals before and after resampling}
\label{fig:repair_freq}
\end{adjustwidth}
\end{figure}

As described in section \ref{rep_freq}, the total number of samples are different for the accelerometer and gyroscope sensors of the smartphone and the smartwatch. This difference was caused by different sampling frequencies used for recording the data. After repairing the frequency problem (i.e., resampling the signals to set all signals at 20Hz), the total number of samples are close to each other for the accelerometer and gyroscope sensor of the smartphone and the smartwatch. Table \ref{tab:length_of_sig} shows the size of each dataset before and after repairing the frequency. Figure \ref{fig:repair_freq} shows an example of resampling the signal. Figure \ref{fig:repair_freq_1} shows the original signal, which was recorded at 50Hz, but based on the description of the dataset, the time was calculated by assuming that the frequency is 20Hz. Figure \ref{fig:repair_freq_2} shows the resampled signal at 20Hz.

\begin{table}[h]
    \centering
    \caption{Total number of samples for each sensor after resampling}
    \label{tab:length_of_sig}
    \resizebox{0.65\columnwidth}{!}{ 

\begin{tabular}{lll}
                                          & \multicolumn{2}{l}{Total samples for all activities}                                 \\ \cline{2-3} 
\multicolumn{1}{l|}{}                     & \multicolumn{1}{l|}{Original signals} & \multicolumn{1}{l|}{Resampled Signals} \\ \hline
\multicolumn{1}{|l|}{Phone accelerometer} & \multicolumn{1}{l|}{4,804,403}        & \multicolumn{1}{l|}{3,261,800}         \\ \hline
\multicolumn{1}{|l|}{Phone gyroscope}     & \multicolumn{1}{l|}{3,608,635}        & \multicolumn{1}{l|}{3,251,188}         \\ \hline
\multicolumn{1}{|l|}{Watch accelerometer} & \multicolumn{1}{l|}{3,777,046}        & \multicolumn{1}{l|}{3,288,129}         \\ \hline
\multicolumn{1}{|l|}{Watch gyroscope}     & \multicolumn{1}{l|}{3,440,342}        & \multicolumn{1}{l|}{3,268,639}         \\ \hline
\end{tabular}
} 
\end{table}

To see the result of repairing the dataset on the classifier's performance, we trained and tested a dCNN model using the original and the repaired WISDM datasets. Table \ref{tab:cross_hardware} shows the result of applying a trained dCNN model on both datasets. Table \ref{tab:cross_hardware_1} shows the micro average F1-score while using the original WISDM dataset and Table  \ref{tab:cross_hardware_2} shows the micro average F1-score while using the repaired WISDM dataset for five activities (i.e., Walking, Jogging, Stairs, Sitting, and Standing). The first row (for both Tables) shows the scenario in which the model was trained on signals from a smartphone accelerometer. The second row (for both Tables) shows the scenario in which the model was trained on signals from a smartwatch accelerometer.  The first column (for both Tables) shows the scenario in which the model was tested on signals from a smartphone accelerometer. The second column (for both Tables) shows the scenario in which the model was tested on signals from a smartwatch accelerometer. 

\begin{table}[t]
\caption{Cross hardware performance: Comparing micro average F1-score of dCNN models applied on the original data and the repaired data for five activities (i.e., Walking, Jogging, Stairs, Sitting, and Standing)}
\label{tab:cross_hardware}
\begin{subtable}[h]{0.35\textwidth}
    \centering
    \caption{Using original signals}
    \label{tab:cross_hardware_1}
    \resizebox{\columnwidth}{!}{
    \begin{tabular}{cccc}
        \multicolumn{2}{c}{} & \multicolumn{2}{c}{\textbf{Test on}}   \\ 
        \cline{3-4}
        \multicolumn{2}{c}{} & \multicolumn{1}{c}{Phone} & \multicolumn{1}{c}{Watch} \\
        \cmidrule(l){2-4} 
        \multirow{2}{*}{\begin{sideways}\textbf{Train on}\end{sideways}} 
        &\multicolumn{1}{c|}{Phone} & \multicolumn{1}{c}{0.76} & \multicolumn{1}{c}{0.73} \\
        &\multicolumn{1}{c|}{Watch} & \multicolumn{1}{c}{0.57} & \multicolumn{1}{c}{0.78} \\
    \end{tabular}
    } 
    \end{subtable}
    \hfill
\begin{subtable}[h]{0.35\textwidth}
    \centering
    \caption{Using repaired signals}
    \label{tab:cross_hardware_2}
    \resizebox{\columnwidth}{!}{
    \begin{tabular}{cccc}
        \multicolumn{2}{c}{} & \multicolumn{2}{c}{\textbf{Test on}}   \\ 
        \cline{3-4}
        \multicolumn{2}{c}{} & \multicolumn{1}{c}{Phone} & \multicolumn{1}{c}{Watch} \\
        \cmidrule(l){2-4} 
        \multirow{2}{*}{\begin{sideways}\textbf{Train on}\end{sideways}} 
        &\multicolumn{1}{c|}{Phone} & \multicolumn{1}{c}{0.79} & \multicolumn{1}{c}{0.78} \\
        &\multicolumn{1}{c|}{Watch} & \multicolumn{1}{c}{0.80} & \multicolumn{1}{c}{0.80} \\
    \end{tabular}
    } 
\end{subtable}
\end{table}

\section{Discussion}
\label{discussion}
We need correct and complete data to train a high-performance machine learning model. The results from training the same model on the two versions of the WISDM dataset (i.e., the original WISDM and the repaired version rWISDM) show a significant improvement in the performance and robustness of the classifier after repairing the dataset, as shown in section \ref{result}.

Table \ref{tab:length_of_sig} shows the size of each dataset before and after repairing the frequency. After repairing the frequency problem (i.e., resampling the signals to set all signals at 20Hz), the total number of samples are close to each other for signals recorded from the accelerometer/gyroscope sensors of the smartphone/smartwatch. The full length of signals from each sensor is almost equal to the number of subjects (51) times the number of activities (17) times the activities' duration (180s) times the frequency (20Hz). The slight difference between datasets' length is that the activities duration was not precisely 180 seconds for all activities and subjects. Also, some of the activities are missing for some of the subjects (e.g., activity B for subject 1609, activities B and F for subject 1616, and activities C and F for subject 1642, in smartphone-accelerometer dataset).

By comparing Table \ref{tab:cross_hardware_1} to Table  \ref{tab:cross_hardware_2}, we see that the micro average F1-score was increased after repairing the dataset.  
Comparing the main diagonal of Table \ref{tab:cross_hardware_1} to the main diagonal of Table  \ref{tab:cross_hardware_2} (which is for classifying five activities and is based on applying the leave-one-out approach), we see that after repairing the data, the micro average F1-score for a model trained and tested on the smartphone data was increased from 0.76 to 0.79. After repairing the data, the micro average F1-score for a model trained and tested on the smartwatch data increased from 0.78 to 0.80. 

Also, we see the enhancement while testing the models for hardware cross-validation. After repairing the data, the micro average F1-score for the model trained on the smartphone data and tested on the smartwatch data was increased from 0.73 to 0.78. 
After repairing the data, the micro average F1-score for the model trained on the smartwatch data and tested on the smartphone data increased from 0.57 to 0.80.

The results show that the model which was trained and tested on the smartphone data has slightly better improvement (i.e., 0.76 to 0.79; improved by 0.03) after repairing the data compared to the model trained and tested on the smartwatch data (i.e., 0.78 to 0.80; improved by 0.02). This could be because the smartwatch had a fixed position on the subjects’ wrists (the exception being left-handed subjects, however, this subjects' meta data was not reported), so it had consistent orientation for the XYZ axes of the sensors. The enhancement for models trained and tested on the smartwatch was primarily because of repairing the frequency issue or could be because of repairing the sensor's axes orientations for left-handed subjects if there is any of them in this dataset.    
\section{Conclusion and Future work}
\label{conclusion}
Many factors are involved in considering a dataset as correct and complete data. These factors are varied based on the underlying circumstance. For example, the voxel size of a medical image such as an MRI or CT scan is significant for classifying tumours based on their volume.

For the correctness of recording human activities using a smartphone, one should clearly define the smartphone position and orientation to ensure consistent sensor data acquisition. Explicitly reporting the free-position and/or free-orientation is a crucial point for the correctness of the dataset if there is no protocol for the position and the orientation of the smartphone. Furthermore, recording the signals with a fixed sampling frequency or reporting the various recording frequencies is essential for creating a correct dataset.  

One of the highlighted strengths of the WISDM dataset is publishing the raw data. It was only because of the raw data that we could find and correct those inconsistencies with the recording. The raw data enable researchers to develop their own preprocessing and manipulating methods, where some datasets such as UCI-HAR \cite{DS_UCI_HAR2013} provided manipulated data (e.g., windows with fixed size and fixed overlap, subtracting the gravity from the total acceleration signals, not reporting the Unix time, etc.) which prevents some investigation and validation of the data. Providing the raw data offers researchers the ability to design new algorithms and methods for preprocessing and manipulating the data.  
We strongly recommend that all researchers who kindly plan to create their public dataset upload all raw data.

We also investigated if the recording for the accelerometer and gyroscope were simultaneous for each subject activity. We found that the recorded Unix time was different for the first sample point of the accelerometer and gyroscope for most of the activities of each subject. As mentioned in section \ref{rep_freq}, the recorded Unix time is not correct considering the experiment year (i.e., 2017). Still, the increase in timestamp is correct and consistent, which was used for calculating the sampling frequency. We recommend more investigation in this topic to use the WISDM accelerometer and gyroscope data together for HAR. 
 
Since our goal was to use only accelerometer data, our investigation focused on signals from the accelerometer of smartphone and smartwatch for five activities (i.e., Walking, Jogging, Stairs, Sitting, Standing). We also found and repaired the same issue with the sampling frequency of recorded gyroscope signals, but did not use it for the classification. We suggest that the interested researchers work on the recorded data from the gyroscope and the data for activities other than those five aforementioned activities.   

\vspace{6pt} 



\authorcontributions{Conceptualization, M.H., T.D.;Formal analysis: M.H.; funding acquisition, T.D.; project administration, T.D.; Writing – original draft, M.H.; Writing – review \& editing,: M.H., T.D. All authors have read and agreed to the published version of the manuscript.}

\funding{This research was funded by the Canadian Department of National Defence Innovation for Defence Excellence and Security (IDEaS) under Award CFPMN2-17.}

\institutionalreview{Not applicable.}

\informedconsent{Not applicable.}

\dataavailability{Not applicable.} 


\conflictsofinterest{The authors declare no conflict of interest.} 



\abbreviations{Abbreviations}{
The following abbreviations are used in this manuscript:\\

\noindent 
\begin{tabular}{@{}ll}
AI & Artificial Intelligence\\
CT & Computed Tomography\\
DL & Deep Learning\\
dCNN &  deep Convolutional Neural Network\\ 
HAR & Human Activity Recognition\\
HPC & High-Performance Computing\\
MRI & Magnetic Resonance Imaging\\
NN & Neural Network\\
RF &  Random Forrest\\
SVM &  Support Vector Machine
\end{tabular}
}

\begin{adjustwidth}{-\extralength}{0cm}

\reftitle{References}



\bibliography{HAR1_bib}

\end{adjustwidth}

\end{document}